\documentclass[twocolumn,showpacs,preprintnumbers,amsmath,amssymb]{revtex4}
\usepackage{graphicx}
\usepackage{dcolumn}
\usepackage{bm}
\usepackage{subfigure}
\usepackage{color}

\begin{document}

\title{Asymptotic behavior of double parton distribution functions}

\author{A.M.~Snigirev}
\affiliation{M.V. Lomonosov Moscow State University, D.V. Skobeltsyn
Institute of Nuclear Physics, 119991, Moscow, Russia }

\date{\today}
\begin{abstract}
The double parton distribution functions are investigated in the region of small
longitudinal momentum fractions in the leading logarithm approximation of perturbative QCD. It is shown that  these functions have the factorization property in the case of one slow and one fast parton. 
\end{abstract}
\pacs{12.38.-t}


\maketitle
\section{\label{sec1}Introduction}
Multiple parton interactions  are one of the most common, yet poorly understood~\cite{mpi10}, phenomena at the LHC. It is usually assumed that only one hard  interaction takes place per hadron-hadron collision together with multiple soft interactions. This assumption is typically justified on the grounds that the probability of a hard parton-parton interaction in a collision is very small. Thus the probability of having two or more hard interactions in a collision is strongly suppressed in comparison with the single interaction probability. Nevertheless hadron collisions in which two (or more) distinct pairs of partons hard scatter are possible.

The presence of such multiple parton interactions in high energy hadron-hadron collisions has been convincingly  demonstrated by the AFS~\cite{AFS}, UA2~\cite{UA2} and CDF~\cite{cdf_4jets} collaborations using events with  the four-jet final state and later by the CDF Collaboration~\cite{cdf} and quite recently by the D0 Collaboration~\cite{D0} using events with the $\gamma+3$ jets final state.  The possibility of observing two separate hard  collisions was proposed long ago and their theoretical study goes back to the early days of the parton model~\cite{landshoff,takagi,goebel} with subsequent extension to perturbative QCD~\cite{paver,humpert,odorico,Mekhfi:1983az, Ametller:1985tp,Halzen:1986ue,sjostrand,Mangano:1988sq,Godbole:1989ti,Drees:1996rw,trelani,trelani2,sjostrand2,sjostrand3,strikman,Gaunt:2009re,Calucci:2010wg,Blok:2010ge,Strikman:2010bg}. 

A greater rate of events containing multiple hard interactions is anticipated at the LHC  with respect to the experiments mentioned above due to the much higher luminosity and greater energy of the LHC. Moreover the products from multiple interactions will represent an important background~\cite{del,DelFabbro:2002pw, Hussein:2006xr, Hussein:2007gj} to signals from the Higgs and other interesting processes and certain types of multiple interactions will have distinctive signature~\cite{Kulesza:1999zh,Cattaruzza:2005nu,maina,Berger:2009cm} facilitating a detailed investigation of these processes experimentally. Therefore it is extremely important to combine theoretical efforts in order to achieve a better description of multiple interactions, in particular, double scattering which will be the dominant multiple scattering mode at the LHC.

With the only assumption of factorization of the two hard parton processes $A$ and $B$ the inclusive cross section of a double parton scattering process in a hadron collision may be written as~\cite{paver, Cattaruzza:2005nu, Gaunt:2009re}
\begin{eqnarray} 
\label{hardAB}
\sigma^D_{(A,B)} = \frac{m}{2} \sum \limits_{i,j,k,l} \int \Gamma_{ij}(x_1, x_2, b; Q^2_1, Q^2_2) \hat{\sigma}^A_{ik}(x_1, x_1^{'})\nonumber\\
\times \hat{\sigma}^B_{jl}(x_2, x_2^{'})
 \Gamma_{kl}(x_1^{'}, x_2^{'}, b; Q^2_1, Q^2_2) dx_1 dx_2 dx_1^{'} dx_2^{'} d^2b. 
\end{eqnarray} 
Here $\Gamma_{ij}(x_1, x_2, b; Q^2_1, Q^2_2)$ are the double parton distribution functions, depending on the longitudinal momentum fractions $x_1$ and $x_2$ and on the relative transverse distance $b$ of the two parton undergoing the hard processes $A$ and $B$ at the scales $Q_1$ and $Q_2$. $\hat{\sigma}^A_{ik}$ and $\hat{\sigma}^B_{jl}$ are the parton-level subprocess cross sections. The factor $m/2$ is a consequence of the symmetry of the expression for interchanging parton species $i$ and $j$. $m=1$ if $A=B$ and $m=2$ otherwise.

It is often assumed that the double parton distribution functions may be decomposed in terms of longitudinal and transverse components as follows:
\begin{eqnarray} 
\label{DxF}
\Gamma_{ij}(x_1, x_2, b; Q^2_1, Q^2_2) = D^{ij}_h(x_1, x_2; Q^2_1, Q^2_2) F_{ij}(b).
\end{eqnarray} 
$F_{ij}(b)$ represents the parton pair density in transverse space and is not calculated in the framework of perturbative QCD. Existing models typically use Gaussian or exponential form to describe the $F_{ij}(b)$, or a sum of Gaussian/exponential terms~\cite{sjostrand2, Bahr:2008dy}.

$D^{ij}_h(x_1, x_2; Q^2_1, Q^2_2)$ is interpreted as the inclusive probability of finding a parton $i$ with the longitudinal momentum traction $x_1$ at scale $Q_1$ and a parton $j$ with the longitudinal momentum traction $x_2$ at scale $Q_2$ in a hadron $h$. The behavior of these distribution functions 
$D^{ij}_h(x_1, x_2; Q^2)=D^{ij}_h(x_1, x_2; Q^2, Q^2)$ with the two hard scale set equal were investigated in Refs.~\cite{Kirschner:1979im,Shelest:1982dg} in the leading logarithm approximation of perturbative QCD. The equations dictating the scaling violations (i.e. $Q^2$ dependence) of the double parton distributions were derived therein. A crucial prediction~\cite{snig03, snig04, snig08} of this is that even if the double parton distribution functions may be taken to be equal to products of single distribution functions at one scale, then at any other scale the double distribution functions will deviate from factorized forms  used often in the current analysis. Quite recently the equations of Refs.~\cite{Kirschner:1979im,Shelest:1982dg} have been numerically integrated~\cite{Gaunt:2009re} to produce a set of publicity available double parton distribution function grids covering the ranges $10^{-6}<x_1<1$, $10^{-6}<x_2<1$, and $1<Q^2<10^9$ GeV$^2$. Thus the correlations induced by QCD evolution may be evaluated already under condition of real experiments and their possible manifestation was discussed in Refs.~\cite{Cattaruzza:2005nu, bandurin, Gaunt:2010pi}. In this case the behavior of double parton distribution functions is extremely important in the kinematical range of relatively small longitudinal momentum fractions which is difficult to investigate numerically with a good accuracy and predictive enough power.

The main purpose of the present paper is to study the asymptotic behavior of double parton distribution functions in the vicinity of the kinematical boundaries analytically. The paper is organized as follows. Section~\ref{sec2}  is devoted to what may be extracted
from perturbative QCD theory on the asymptotic behavior of the two-parton distribution functions at the parton level. The possible phenomenological issues for parton distributions in a hadron are discussed in Sec.~\ref{sec3}. We summarize and conclude in Sec.~\ref{sec4}.

\maketitle
\section{\label{sec2} Double parton distributions in the leading logarithm approximation at parton level}
The analysis~\cite{gribov, lipatov, dokshitzer} of hard processes, $ep$ scattering and
$e^+e^-$ annihilation, in the leading logarithm approximation of perturbative QCD admits a simple interpretation~\cite{lipatov} in the framework of the parton model with a variable cutoff parameter $\Lambda~\sim~ Q$ with respect to the transverse momenta. The dependence of the multiparton distribution and fragmentation functions on the value of this cutoff parameter is determined by the evolution equations. Lipatov suggested a general elegant method~\cite{lipatov} of obtaining these equations in any renormalizable quantum field theory. As an evolution variable one uses the value of hard scale (the transfer momentum squared $Q^2$, most commonly) or its logarithm $\xi=\ln(Q^2/\mu^2)$ or double logarithm 
\begin{equation}
t = \frac{1}{2\pi \beta} \ln \Bigg[1 + \frac{g^2(\mu^2)}{4\pi}\beta
\ln\Bigg(\frac{Q^2}{\mu^2}\Bigg)\Bigg] = \frac{1}{2\pi \beta}\ln\Bigg
[\frac{\ln(\frac{Q^2}{\Lambda^2_{\rm QCD} })}
{\ln(\frac{\mu^2}{\Lambda^2_{\rm QCD}})}\Bigg],
\end{equation}
where $\beta = (33-2n_f)/12\pi$ ${\rm {in~ QCD}},$
$g(\mu^2)$ is the running coupling constant at the reference scale $\mu^2$, $n_f$ is the number of active flavors, $\Lambda_{\rm QCD}$ is the dimensional QCD parameter.

With the evolution variable $t$ (taking into account the behavior of the running coupling constant in one loop approximation explicitly) the DGLAP equations~\cite{gribov,lipatov, dokshitzer, altarelli} for single distributions read more easily
\begin{equation}
\label{e1singl}
 \frac{dD_i^j(x,t)}{dt} = 
\sum\limits_{j{'}} \int \limits_x^1
\frac{dx{'}}{x{'}}D_i^{j{'}}(x{'},t)P_{j{'}\to j}\Bigg(\frac{x}{x{'}}\Bigg).
\end{equation}
\noindent
They describe the scaling violation of the parton distributions $ D^j_i(x,t)$ inside a dressed quark or gluon ($i,j = q/g$). It is interesting that  expression for the kernels $P$ in Lipatov method already includes a regularization at $x \rightarrow x{'}$, which was introduced in Ref.~\cite{altarelli} afterwards based on the momentum conservation.

This method  allows one to obtain also the generalized DGLAP equations for two-parton distributions $D_i^{j_1j_2}(x_1,x_2,t)$, representing the probability that in a dressed
constituent $i$ one finds two bare partons  of  types $j_1$ and $j_2$ with the given
momentum fractions $x_1$ and $x_2$, namely (see Refs.~\cite{Kirschner:1979im, Shelest:1982dg} for details):
\begin{eqnarray}
\label{edouble}
& &\frac{dD_i^{j_1j_2}(x_1,x_2,t)}{dt} \\
& &=\sum\limits_{j_1{'}}
\int\limits_{x_1}^{1-x_2}\frac{dx_1{'}}{x_1{'}}D_i^{j_1{'}j_2}(x_1{'},x_2,t)
P_{j_1{'}
\to j_1} \Bigg(\frac{x_1}{x_1{'}}\Bigg) \nonumber\\
& &+ \sum\limits_{j_2{'}}\int\limits_{x_2}^{1-x_1}
\frac{dx_2{'}}{x_2{'}}D_i^{j_1j_2{'}}(x_1,x_2{'},t)P_{j_2{'} \to j_2}
\Bigg(\frac{x_2}{x_2{'}}\Bigg) \nonumber\\
& &+ \sum\limits_{j{'}}D_i^{j{'}}(x_1+x_2,t) \frac{1}{x_1+x_2}P_{j{'} \to
j_1j_2}\Bigg(\frac{x_1}{x_1+x_2}\Bigg),\nonumber
\end{eqnarray}
where the kernel
\begin{equation}
\frac{1}{x_1+x_2} P_{j{'} \to
j_1j_2}(\frac{x_1}{x_1+x_2})
\end{equation}
is defined without $\delta$-function regularization. 

It is readily  verified by  direct substitution that the solution of Eq.~(\ref{edouble}) can be written via the convolution of single distributions~\cite{Kirschner:1979im, Shelest:1982dg}
\begin{eqnarray}
\label{solution}
& & D_i^{j_1j_2}(x_1,x_2,t) \\
& &= \sum\limits_{j{'}j_1{'}j_2{'}} \int\limits_{0}^{t}dt{'}
\int\limits_{x_1}^{1-x_2}\frac{dz_1}{z_1}
\int\limits_{x_2}^{1-z_1}\frac{dz_2}{z_2}~
D_i^{j{'}}(z_1+z_2,t{'})\frac{1}{z_1+z_2} \nonumber\\
& & \times P_{j{'} \to
j_1{'}j_2{'}}\Bigg(\frac{z_1}{z_1+z_2}\Bigg) D_{j_1{'}}^{j_1}(\frac{x_1}{z_1},t-t{'}) 
D_{j_2{'}}^{j_2}(\frac{x_2}{z_2},t-t{'}).\nonumber
\end{eqnarray}
\noindent 
This coincides with the jet calculus rules~\cite{konishi} proposed originally for the fragmentation functions and is the  generalization of the well-known Gribov-Lipatov relation installed for single functions~\cite{gribov, lipatov, dokshitzer} (the distribution of bare partons inside a dressed constituent  is identical to the distribution of dressed constituents in the fragmentation of a bare parton in the leading logarithm approximation). Nevertheless in a phenomenological employment the direct numerical integration of Eq.~(\ref{edouble}) occurs more effective~\cite{Gaunt:2009re} than the explicit solution~(\ref{solution}) because of the singular $\delta$-like initial conditions of single distributions $D_i^j(x,t)$ (the Green's functions) incoming in this solution.

Eqs.~(\ref{e1singl}) and (\ref{edouble}) are explicitly solved by introducing the Mellin transforms
\begin{eqnarray}
\label{mellin}
 & M_i^{j}(n,t) = \int\limits_{0}^{1}dx x^n~D_{h}^{j}(x,t),\\
 &M_i^{j_1 j_2}(n_1,n_2,t) \nonumber\\
&= \int\limits_{0}^{1}dx_1 dx_2 \theta(1-x_1-x_2)x_1^{n_1}x_2^{n_2}D_{h}^{j_1 j_2}(x_1,x_2,t)
\end{eqnarray}
which lead to a system of ordinary linear-differential equations at the first order:
\begin{equation}
\label{1.19}
dM_i^j (n,t)/dt = \sum\limits_{j{'}} M_i^{j{'}}(n,t) P_{j{'} \to j}(n), 
\end{equation}
\begin{eqnarray}
\label{1.17}
& &dM_i^{j_1j_2}(n_1,n_2,t)/dt \nonumber \\
& &=\sum\limits_{j_1{'}} M_i^{j_1{'}j_2} (n_1,n_2,t)
P_{j_1{'} \to j_1} (n_1) \nonumber\\
& &+ \sum\limits_{j_2{'}} M_i^{j_1j_2{'}}(n_1,n_2,t) P_{j_2{'} \to j_2}(n_2) 
\nonumber \\
& &+ \sum\limits_{j{'}} M_i^{j{'}} (n_1+n_2,t)P_{j{'} \to j_1j_2}(n_1,n_2), 
\end{eqnarray}
where
\begin{eqnarray}
\label{1.21}
P_{i \to i_1}(n)  =  \int\limits_0^1 x^n P_{i \to i_1} (x)dx, \\
P_{i \to i_1i_2}(n_1,n_2)  = \int\limits_0^1 x^{n_1} (1-x)^{n_2} P_{i \to
i_1i_2} (x) dx.   
\end{eqnarray}

In order to obtain the distributions in $x$ representation an inverse Mellin transformation must be performed
\begin{eqnarray}
\label{mellin in}
x D_i^{j}(x,t) =~\int\frac {dn}{2\pi i} x^{-n}~M_{i}^{j}(n,t),
\end{eqnarray}
where the integration runs along the imaginary axis to the right from all
$n$ singularities. This can be done  in a general case only numerically. However the asymptotic behavior can be estimated in some interesting and particularly simple limits in the technique under consideration.

\maketitle
\subsection{\label{finite x} Spectra at finite $x$}

At first let us look at single parton distribution functions near $x=1$. These can be obtained through the study of large moments (which clearly emphasize $x\sim 1$). One obtains behaviors of the type~\cite{dokshitzer, konishi}:
\begin{eqnarray}
\label{single x=1}
D_i^{j}(x,t)\big|_{x \to 1} \sim C_i^j(x,t)(1-x)^{2C_Ft+b_{ij}},
\end{eqnarray}
where $C_F=(N_c^2-1)/2N_c=4/3, N_c=3$ and $C_i^j(x,t)$ depend logarithmically upon $(1-x)$ and in a complicated but known way upon $t$, whereas $b_{ij}$ are just some integer numbers.
The values of $C_i^j(x,t), b_{ij}$ are given in Table 4 of Ref.~\cite{konishi}(NP) for all cases. For illustration we write here the asymptotic expression for the distribution of
valence quarks (the probability of finding in a quark a quark of the same species) at $1-x \ll 1$~\cite{dokshitzer, konishi}: 
\begin{eqnarray}
\label{valence}
D^{\rm val}(x,t)=\exp{((3/2-2\gamma_E)C_F t)}\frac{(1-x)^{2C_Ft-1}}{\Gamma(2C_Ft)},
\end{eqnarray} 
where $\gamma_E=0.5772$... is the Euler's constant. Indeed, at large n the kernel 
\begin{eqnarray}
\label{kernel}
P_{q\to q}(n)\simeq (3/2-2\gamma_E-2\ln{n})C_F,  
\end{eqnarray} 
and an inverse Mellin transformation (\ref{mellin in}) can be performed analytically  using the exact solution 
\begin{eqnarray}
\label{valence n}
M^{\rm val}(n,t) =\exp{(P_{q\to q}(n) t)}
\end{eqnarray} 
in $n$ representation. The characteristic moment being significant in integral (\ref{mellin in}) is defined by the saddle point equation
\begin{eqnarray}
\label{saddle eq} 
n_0 = 2C_F t/(1-x),~~~~ n_0 \gg 1.
\end{eqnarray}

For the two-parton distribution functions the limit $x_1+x_2 \to 1$ can be investigated by looking into large moments:
\begin{eqnarray}
\label{mellindouble}
\int\limits_{0}^{1}dx_1 dx_2 (x_1+x_2)^{n}x_1x_2D_{i}^{j_1 j_2}(x_1,x_2,t).
\end{eqnarray}
The result is a behavior of the type~\cite{konishi}:
\begin{eqnarray}
\label{double x1+x2=1}
D_i^{j_1j_2}(x_1,x_2,t) \sim A_i^{j_1 j_2}(1-x_1-x_2)^{2C_Ft+\delta_{i,j_1j_2}},
\end{eqnarray}
where $A_i^{j_1 j_2}$ have at most a logarithmic dependence on $(1-x_1-x_2)$. The exponents $\delta_{i,j_1j_2}$ are some integer numbers given in Table 5 of Ref.~\cite{konishi}(NP).

The analog of a double-Regge limit 
\begin{eqnarray}
\label{regge}
1-x_1 \ll 1,~~~ 1-x_1-x_2\ll 1-x_1
\end{eqnarray}
can be done by studying double moments $M_i^{j_1 j_2}(n_1,n_2,t )$ for $n_1\gg n_2 \gg 1$. By inverting the Mellin transform one then finds
\begin{eqnarray}
\label{double regge}
& &D_i^{j_1j_2}(x_1, x_2,t)  \\
& &\sim H(t)(1-x_1)^{k_{i,j_1j_2}}(1-x_1-x_2)^{2C_Ft+h_{i,j_1j_2}},\nonumber
\end{eqnarray}
up to logarithmic terms. The exponents $k_{i,j_1j_2}$ and $h_{i,j_1j_2}$ can be computed and are given in Table 6 of Ref.~\cite{konishi}(NP).

The obtained asymptotic behaviors~(\ref{single x=1}, \ref{double x1+x2=1}, \ref{double regge}) admit an amusing analogy with Regge theory as it was pointed out in Ref.~\cite{konishi}.
This helps one also to understand better why the phase space factor in the improved ansatz~\cite{Gaunt:2009re} for the initial conditions of parton distributions at the hadron level must include the factors $(1-x_1-x_2), (1-x_1), (1-x_2)$ with the exponents depending on parton types in order to satisfy the momentum and number sum rules.

\maketitle
\subsection{\label{small x} Spectra at small $x$}
The increase in the average number of decays of each parton with increasing $t$ leads, as can be readily understood, to a large increase in the number of fields, gluons and sea quarks, with small longitudinal momentum fractions. Technically, integral~(\ref{mellin in})
when $x\ll1$ is determined by the first singularity with respect to the n of the moments
$M_i^j(n,t)$, which is related to the pole of the larger eigenvalue $\Lambda_{+}(n)$ of the system~(\ref{1.19}) when $n=0$:
\begin{eqnarray}
\label{lambda}
\Lambda_{+}(n) \simeq \frac{2N_c}{n}-a+..., ~~~~~n\ll1,
\end{eqnarray}
where $a=\frac{11}{6}N_c +\frac{1}{3}n_f(1-2C_F/N_c)$. By inverting the Mellin transform with the approximation~(\ref{lambda})  one then obtains the probability of finding a gluon in a gluon:
\begin{eqnarray}
\label{mellin in x=0}
& & x D_g^{g}(x,t) =~\int\frac {dn}{2\pi i} \exp{[n\ln{(1/x)}+2N_c t/n-at]}\nonumber\\
& &= 4N_ct \exp{[-at]}I_1(v)/v \nonumber\\
& &\simeq 4N_ct v^{-3/2}\exp{[v-at]}/\sqrt{2\pi},
\end{eqnarray}
where $v=\sqrt{8N_ct\ln{(1/x)}}$ and $I_1$ is the usual modified Bessel function. The last approximate equality in Eq.~(\ref{mellin in x=0}) is obtained  by the saddle point method with the saddle point equation 
\begin{eqnarray}
\label{saddle eq2} 
n_0 = \sqrt{2N_c t/\ln{(1/x)}},~~~~ n_0 \ll1.
\end{eqnarray}
The probabilities of finding any other partons in a gluon or a quark are evaluated in the regime of small $x$ by a similar way and can be found in Refs.~\cite{dokshitzer, konishi}.

Due to the generalized Gribov-Lipatov relation mentioned above in the limit of one slow ($x_1\sim0$) and one fast ($x_2=$ finite) parton the two-parton distribution functions become uncorrelated the same as the two-parton fragmentation functions for which this property was established in Ref.~\cite{konishi}. The factorization property is significant in a phenomenological employment and its proof is worth to be reviewed once more because of a very old affair.

The exact solution for single distributions in the moments representation can be written symbolically in a matrix form
\begin{eqnarray}
\label{solution n}
M_i^{j}(n,t) =(\exp{P(n) t})_i^j
\end{eqnarray}
The solution for the two-parton distribution functions in this representation reads
\begin{eqnarray}
\label{1.22}
& &M_i^{j_1j_2}(n_1,n_2,t)  \nonumber\\ 
& & =\sum\limits_{jj_1{'}j_2{'}} \int\limits_0^t dt{'}
(\exp{P(n_1+n_2)t^{'}})_i^j [P_{j \to j_1{'}j_2{'}}(n_1,n_2)]  \nonumber \\
& &\times (\exp{P(n_1)(t-t{'})})_{j_1{'}}^{j_1} (\exp{P(n_2)(t-t{'})})_{j_2{'}}^{j_2}.
\end{eqnarray}
Going further through a Laplace transform to the variable $E$ one finds
\begin{eqnarray}
\label{laplace}
& &M_i^{j_1 j_2}(n_1,n_2,E) 
= \int\limits_{0}^{\infty}dt e^{-Et}M_i^{j_1 j_2}(n_1,n_2,t)\nonumber\\
& & =\sum\limits_{jj_1{'}j_2{'}} 
(\frac{1}{E-P(n_1+n_2)})_i^j [P_{j \to j_1{'}j_2{'}}(n_1,n_2)]  \nonumber \\
& &\times (\frac{1}{E-P(n_1)-P(n_2)})_{j_1{'} j_2{'} }^{j_1 j_2}.
\end{eqnarray}
Let us consider now the correlation function and its moments:
\begin{eqnarray}
\label{corr}
& &C_i^{j_1j_2}(x_1,x_2,t)\nonumber\\
& & =D_i^{j_1j_2}(x_1,x_2,t)-D_i^{j_1}(x_1,t)D_i^{j_2}(x_2,t),
\end{eqnarray}
\begin{eqnarray}
\label{laplace corr}
& &C_i^{j_1 j_2}(\epsilon,n,E) \\
& & =\sum\limits_{jj_1{'}j_2{'}} 
(\frac{1}{E-P(\epsilon+n)})_i^j [P_{j \to j_1{'}j_2{'}}(\epsilon,n)]  \nonumber \\
& &\times (\frac{1}{E-P(\epsilon)-P(n)})_{j_1{'} j_2{'} }^{j_1 j_2}
-(\frac{1}{E-P(\epsilon)-P(n)})_{i~i}^{j_1 j_2},\nonumber
\end{eqnarray}
where $\epsilon=0^{+}$ in the moment relative to the slow parton and $n=O(1)$. In Eq.~(\ref{laplace corr}) various factors are divergent since both $P(\epsilon)$ and $P_{j \to j_1{'}j_2{'}}(\epsilon,n)$ contain divergent matrix elements. Since, however,
\begin{eqnarray}
\label{limit epsilon}
P_{j \to j_1{'}j_2{'}}(\epsilon,n)\sim P_{j j_1^{'}}(\epsilon)~~~ P(\epsilon+n)\sim P(n)+ O(\epsilon)
\end{eqnarray}
one finds that $C_i^{j_1 j_2}(\epsilon,n,E)$ is NOT divergent. In other words the divergent part in $D_i^{j_1j_2}(x_1,x_2,t)$ at $x_1\to 0$ and finite $x_2$ factorizes into $D_i^{j_1}(x_1,t)$ times $D_i^{j_2}(x_2,t)$. This result can be generalized to $m$-parton distribution functions, i.e., when one of the partons becomes wee its divergence factors out.
\maketitle
\section{\label{sec3} Double parton distributions in hadrons}
Of course, it is interesting to find out the phenomenological issue of this
parton level consideration. This can be done within the well-known
factorization of soft and hard stages (physics of short and long
distances)~\cite{collins}. As a result the equations (\ref{e1singl}, \ref{edouble}, \ref{1.19}, \ref{1.17}) 
 describe the evolution of parton distributions in a hadron with
$t ~(Q^2)$, if one replaces the index $i$ by index $h$ only. However, the initial
conditions for new equations at $t=0 ~(Q^2=\mu^2)$ are unknown a priori and must
be introduced phenomenologically or must be extracted from experiments 
or some models dealing with physics of long distances [at the parton level: 
$D_{i}^{j}(x,t=0)~= ~\delta_{ij} \delta(x-1)$; ~$D_i^{j_1j_2}(x_1,x_2,t=0)~=~0$].
Nevertheless the solutions of the generalized  
DGLAP evolution equations with the given initial
conditions may be written as before via the convolution of single distributions and in the moments representation they read
\begin{eqnarray}
\label{1.37}
M_h^{j_1j_2}(n_1,n_2,t)  =  M_{h0}^{j_1j_2}(n_1,n_2,t) \nonumber \\ 
+ \sum\limits_{j_1{'}j_2{'}} M_h^{j_1{'}j_2{'}}(n_1,n_2,0) M_{j_1{'}}^{j_1}(n_1,t) M_{j_2{'}}^{j_2}(n_2,t), 
\end{eqnarray}
where
\begin{eqnarray}
\label{1.36}
& &M_{h0}^{j_1j_2}(n_1,n_2,t) = \nonumber\\
& &\sum\limits_{i} 
M_h^i (n_1+n_2,0) M_{i}^{j_1j_2}(n_1,n_2,t) 
\end{eqnarray} 
are the particular solutions of the complete equations with the zero initial conditions at the hadron level and
\begin{eqnarray}
\label{1.36q}
M_{i}^{j_1j_2}(n_1,n_2,t) = \sum\limits_{jj_1{'}j_2{'}} \int\limits_0^t dt{'}
M_i^j (n_1+n_2,t{'})\nonumber \\
P_{j \to j_1{'}j_2{'}}(n_1,n_2) 
 M_{j_1{'}}^{j_1}(n_1,t-t{'}) M_{j_2{'}}^{j_2}(n_2,t-t{'})
\end{eqnarray}
are the particular solutions of the complete equations with the zero initial conditions at the parton level. While the second term in the expression~(\ref{1.37}) represents the solutions
of the homogeneous evolution equations with the given initial conditions
$M_h^{j_1{'}j_2{'}}(n_1,n_2,0)$. These nonperturbative unknown two-parton initial conditions are just the reckoning for the unsolved confinement problem. If one assumes that in the moments representation there is the approximate factorization of these initial conditions
\begin{eqnarray}
\label{fact in}
M_h^{j_1{'}j_2{'}}(n_1,n_2,0)\simeq M_h^{j_1{'}}(n_1,0) M_h^{j_2{'}}(n_2,0),
\end{eqnarray}
then the solutions of the homogeneous evolution equations will be factorized in x representation.

The exact solutions~(\ref{1.37}) in the moments representation together with some additional reasonable assumptions on the initial conditions allow one to extract the asymptotic behavior of the distribution functions already in hadrons near the kinematical boundaries as it has be done for the functions of the parton level. Since the kinematical range of relatively small longitudinal momentum fractions is extremely important under condition of real experiments we focus on the asymptotic behavior of two-parton distribution functions just in this range. 

However some interesting results at finite $x$  can be found in Ref.~\cite{vendramin} for two-particle fragmentation functions  and can be extended to two-parton distribution functions taking into account the difference between the solutions of homogeneous evolution equations for the fragmentation and distribution functions~\cite{snig2}. Thus the asymptotic behaviors like (\ref{double x1+x2=1}, \ref{double regge}) can be obtained at the hadron level also assuming that the initial conditions can be evaluated in the saddle points which were defined at the parton level or/and fixing  the asymptotic form  of initial conditions near the kinematical boundary. 

In the important limit of one slow ($x_1\sim0$) and one fast ($x_2=$ finite) parton the two-parton distribution functions are determined by the first singularity with respect to the $n_1$ of the moments $M_h^{j_1j_2}(n_1,n_2,t)$ with $n_2=O(1)$. In this case in accordance with the consideration above the perturbatively calculated particular solutions~(\ref{1.36q}) of the parton level factorizes into $M_i^{j_1}(n_1,t)$ times $M_i^{j_2}(n_2,t)$. Thereafter the general solutions~(\ref{1.37}) read
\begin{eqnarray}
\label{1.37fact}
& & M_h^{j_1j_2}(n_1,n_2,t)  \\
& & =\sum\limits_{i} 
M_h^i (n_1+n_2,0) M_i^{j_1}(n_1,t) M_i^{j_2}(n_2,t) \nonumber\\ 
& & + \sum\limits_{j_1{'}j_2{'}} M_h^{j_1{'}}(n_1,0) M_h^{j_2{'}}(n_2,0) M_{j_1{'}}^{j_1}(n_1,t) M_{j_2{'}}^{j_2}(n_2,t), \nonumber
\end{eqnarray}
the natural approximate factorization of the initial conditions in the form~(\ref{fact in}) being supposed.

Since in the limit $\epsilon \to 0^+$~\cite{dokshitzer, konishi}
\begin{eqnarray}
\label{solution epsilon}
& &M_g^{g}(\epsilon,t) = \frac{N_c}{C_F}M_q^{g}(\epsilon,t) =\exp{(\Lambda_{+}(\epsilon) t)},\\
& &M_g^{q}(\epsilon,t) = \frac{N_c}{C_F}M_q^{q}(\epsilon,t) =\frac{2n_f\epsilon}{6N_c}\exp{(\Lambda_{+}(\epsilon)t)},
\end{eqnarray}
then the ratio of the particular solutions (correlations induced by evolution) to the solutions of homogeneous equations (giving approximate factorization) reduces to
\begin{eqnarray}
\label{ratio}
& & \frac{M_{h0}^{g j_2}(\epsilon,n_2,t)}{M_{h\rm fact}^{g j_2}(\epsilon,n_2,t)}  \\
& & =\frac {\sum\limits_{i} 
M_h^i (\epsilon+n_2,0) M_i^{g}(\epsilon,t) M_i^{j_2}(n_2,t)} { \sum\limits_{j_1{'}} M_h^{j_1{'}}(\epsilon,0) M_{j_1{'}}^{g}(\epsilon,t) \sum\limits_{j_2{'}} M_h^{j_2{'}}(n_2,0)  M_{j_2{'}}^{j_2}(n_2,t)} \nonumber\\
& & =\frac {M_h^g(n_2,0)M_g^{j_2}(n_2,t)+ \frac{C_F}{N_c}\sum\limits_{q} 
M_h^q (n_2,0)  M_q^{j_2}(n_2,t)} { [M_h^g(\epsilon,0)+\frac{C_F}{N_c}\sum\limits_{q} M_h^{q}(\epsilon,0) ]  M_{h}^{j_2}(n_2,t)} \nonumber
\end{eqnarray}
and the universal divergent factor $\exp{(\Lambda_{+}(\epsilon) t)}$ is cancelled out. One should note that this cancellation takes place also in the general case without the supposed factorization of initial conditions in the form~(\ref{fact in}). Therefore the ratio~(\ref{ratio}) admits the finite limit at $\epsilon \to 0^{+}$ provided that the initial number of quarks and gluons  in a hadron is finite, i.e. $M_h^g(0,0)$ and $M_h^q(0,0)$ are not singular and can be evaluated in the saddle point~(\ref{saddle eq2}). Without a factor $C_F/N_c$ before the quark sum the numerator in the ratio under consideration~(\ref{ratio}) is simply equal to $M_h^{j_2}(n_2,t)$. Therefore the suppression of correlations induced by evolution with respect to the factorization component will be mainly determined by the gluon and quark multiplicities in the initial conditions (with a possible correction related to the estimation of zero moments in the the saddle point): 
\begin{eqnarray}
\label{x ratio}
& & \frac{D_{h0}^{g j_2}(x_1, x_2,t)}{D_{h\rm fact}^{g j_2}(x_1,x_2,t)} \Big|_{x_1 \to 0} \nonumber\\
& &\sim \frac {1} { M_h^g(0,0)+\frac{C_F}{N_c}\sum\limits_{q} M_h^{q}(0,0) }. 
\end{eqnarray}
The similar result is obtained if  one slow quark is considered instead of a slow gluon.

Thus in the important limit of one slow ($x_1\sim0$) and one fast ($x_2=$ finite) parton the two-parton distribution functions possess practically the factorization property. The additional contribution induced by evolution being suppressed by the initial gluon and quark multiplicities in comparison with the ``genuine'' factorization component (the solution of homogeneous equation).

\maketitle
\section{\label{sec4} Conclusions}
The asymptotic behavior of the two-parton distribution functions has been investigated near  the kinematical boundaries. At $x$ close to 1 these functions include the factors 
$(1-x_1-x_2), (1-x_1), (1-x_2)$ with the exponents depending on parton types. These exponents are known at the parton level and can be calculated in principle at the hadron level fixing  the asymptotic form  of initial conditions near this kinematical boundary. The importance of the factors under consideration was found in Ref.~\cite{Gaunt:2009re} at the numerical integration of evolution equations in $x$ representation taking into account the momentum and number sum rules. 

The two-parton distribution functions become practically uncorrelated in the kinematical range of relatively small longitudinal momentum fractions which is extremely important under condition of real experiments. The additional ``factorization'' contribution induced by evolution being suppressed by the initial gluon and quark multiplicities in comparison with the ``genuine'' factorization component (the solution of homogeneous equation) in the case of
one slow ($x_1\sim0$) and one fast ($x_2=$ finite) parton. This revealed property explains
why the indications from the experimental observation of double parton scattering at CDF~\cite{cdf} are not in favor of correlation effects in longitudinal momentum fractions.
It is extremely difficult to select events experimentally with double parton scattering in which both partons are enough fast and therefore are strongly correlated in longitudinal momentum fractions. In this connection it is interesting to study the double parton distribution functions beyond the leading logarithm approximation over $Q^2$ (for instance in the BFKL regime~\cite{bfkl, bfkl2, bal, ryskin} $Q^2 =\rm {const}$ and $\ln{(1/x)} \to \infty $ with the possible nontrivial nonfactorization behavior of the two-parton distribution functions just at small $x$).
\begin{acknowledgments}
Discussions with D.V.~Bandurin, E.E.~Boos, A.V.~Leonidov, S.V.~Molodtsov, G.M.~Zinovjev and N.P.~Zotov are gratefully acknowledged. 
This work is partly supported by Russian Foundation for Basic Research Grants No. 08-02-91001, No. 08-02-92496 and No. 10-02-93118, the President of Russian Federation for support of Leading Scientific Schools Grant No. 1456.2008.2.
\end{acknowledgments}


\end{document}